\documentclass[12pt]{iopart}
\usepackage{graphicx}
\usepackage{iopams}
\usepackage{amssymb}
\usepackage{multirow}
\usepackage{xcolor}

\newcommand{\ped}[1]{\ensuremath{_{\rm #1}}}

\begin{document}
\title{Mathematical and physical properties of three bands $s\pm$ Eliashberg theory for iron pinictides}
\author{G.A. Ummarino}
\ead{giovanni.ummarino@polito.it}
\address{Istituto di Ingegneria e Fisica dei Materiali, Dipartimento di Scienza Applicata e Tecnologia, Politecnico di
Torino, Corso Duca degli Abruzzi 24, 10129 Torino, Italy; National Research Nuclear University MEPhI (Moscow Engineering Physics Institute), Kashirskoe shosse 31, Moscow 115409, Russia}

\begin{abstract}
The phenomenology of iron pnictide superconductor can be described by a three bands $s\pm$ Eliashberg theory where the mechanism of superconducting coupling is mediated by antiferromagnetic spin fluctuations and whose characteristic energy $\Omega_{0}$ scales with $T_{c}$ according to the empirical law $\Omega_{0} = 4.65k_{B}T_{c}$. This model presents universal characteristics that are independent of the critical temperature such as the link between the two free parameters $\lambda_{13}$ and $\lambda_{23}$ and the ratios $\Delta_{i}/k_{B}T_{c}$.
\end{abstract}
\maketitle
\section{Introduction}
The superconductive compounds based on iron and arsenic have been discovered for more than fifteen years and all experimental data have been successfully reproduced by using the multiband Eliashberg theory. The mechanism responsible for the pairing is mainly due to antiferromagnetic spin fluctuactions. The various compounds can be described mainly by three \cite{Tors1,Tors2,Tors3,Tors4,Tors5}, four \cite{Umma4b} or five \cite{Umma5b} bands models while the two bands model is purely phenomenological and the values of the electron-boson coupling constants have no physical significance. In most cases, the three-band model is sufficient to describe the experimental data relating to these materials. So we will consider the properties of the three-band model where a fundamental role will be played by the assumption that the representative energy of these systems $\Omega_{0}$ is related to the critical temperature by a universal linear relationship \cite{Paglione1,Paglione2} $\Omega_{0} = 4.65k_{B}T_{c}$ and the symmetry of the order parameter is $s\pm$ \cite{Mazin1,Mazin2,Mazin3}.
In the past J.M. Coombes and J.P. Carbotte \cite{Coombes1,Coombes2,Carbi} found that, if the whole energy scales of the electron-phonon spectral function, in the single band s-wave Eliashberg equations, are shrinked or expanded, the rate between the gap and critical temperature doesn't change. This result is exact if the Coulomb pseudopotential is zero.
Examining the Eliashberg equations for a multiband system it is possible to see that this scaling theorem continues to hold and the values of the gaps and of the critical temperature have increased or decreased by the same factor with which the energy scale has increased or decreased. In fact, Eliashberg's equations for a multiband system are the sum of individual pieces where, in each of which, we can expand or restrict the energy scale. Also in this case the result is correct only if all the values of the Coulomb pseudopotential are zero.
The novelty of our work lies, essentially, in shedding light on the universal bond that exists between the coupling constants relating to the single bands.

\section {The Model}
The simplest model to describe the phenomenology of iron pinictides within Eliashberg's theory consists of a $s\pm$ three bands model, two holonic and one electronic. In this way the two gaps of the hole bands, $\Delta_{1}$ and $\Delta_{2}$, have opposite sign to the gap residing on the electron band i.e. $\Delta_{3}$. The interband coupling between hole and electron bands is mainly provided by antiferromagnetic spin fluctuations (\textit{sf}), while phonons can be responsible for the intraband coupling (\textit{ph})~\cite{Mazin1}. The antiferromagnetic spin fluctuation coupling between bands with the same type of charge carriers (holes with holes and electrons with electrons) is zero while the total phonon coupling is very small ($< 0.35$) \cite{Boeri}.
We assume that the symmetry of this system is $s\pm$ and the electron-boson coupling is from antiferromagnetic spin fluctuactions and in very small part from phonons. The interbands constants coupling in the paper are just relative to antiferromagnetic spin fluctuactuions and are positive because we, as usually, change the sign inside the equations.
To calculate the gaps and the critical temperature we use the $s\pm $ wave three-band Eliashberg equations with infinite bandwidth \cite{Eliashberg,Eliashberg1,Chubukov}.
The infinite bandwidth approximation is always applicable in iron pnictides. Eliashberg's equations without this approximation are more complicated \cite{Eliashbergband1,Eliashbergband2} and the solutions diverge appreciably only in striking cases such as for example in strontium titanate ($SrTiO$) \cite{SrTiO}. In the case of one-band systems, the same results are obtained except in extreme cases where the width of the conduction band is comparable with the phonon energies \cite{Eliashbergband}. We have to solve six coupled equations: three for the gaps $\Delta_{i}(i\omega_{n})$ and three for the renormalization functions $Z_{i}(i\omega_{n})$, where $i$ is a band index (that ranges between $1$ and $3$) and $\omega_{n}$ are the Matsubara frequencies. The imaginary-axis equations \cite{Umma1,Umma2,Umma3}, when the Migdal theorem \cite{Migdal} is valid, read:
\begin{eqnarray}
&&\omega_{n}Z_{i}(i\omega_{n})=\omega_{n}+ \pi T\sum_{m,j}\Lambda^{Z}_{ij}(i\omega_{n},i\omega_{m})N^{Z}_{j}(i\omega_{m})+\nonumber\\
&&+\sum_{j}\big[\Gamma^{N}\ped{ij}+\Gamma^{M}\ped{ij}\big]N^{Z}_{j}(i\omega_{n})
\label{eq:EE1}
\end{eqnarray}
\begin{eqnarray}
&&Z_{i}(i\omega_{n})\Delta_{i}(i\omega_{n})=\pi
T\sum_{m,j}\big[\Lambda^{\Delta}_{ij}(i\omega_{n},i\omega_{m})-\mu^{*}_{ij}(\omega_{c})\big]\times\nonumber\\
&&\times\Theta(\omega_{c}-|\omega_{m}|)N^{\Delta}_{j}(i\omega_{m})
+\sum_{j}[\Gamma^{N}\ped{ij}-\Gamma^{M}\ped{ij}]N^{\Delta}_{j}(i\omega_{n})\phantom{aaaaaa}
 \label{eq:EE2}
\end{eqnarray}
where $\Gamma^{N}\ped{ij}$ and $\Gamma^{M}\ped{ij}$ are the scattering rates from non-magnetic and magnetic impurities, $\Lambda^{Z}_{ij}(i\omega_{n},i\omega_{m})=\Lambda^{ph}_{ij}(i\omega_{n},i\omega_{m})+\Lambda^{sf}_{ij}(i\omega_{n},i\omega_{m})$ and
$\Lambda^{\Delta}_{ij}(i\omega_{n},i\omega_{m})=\Lambda^{ph}_{ij}(i\omega_{n},i\omega_{m})-\Lambda^{sf}_{ij}(i\omega_{n},i\omega_{m})$
where
\[\Lambda^{ph,sf}_{ij}(i\omega_{n},i\omega_{m})=2
\int_{0}^{+\infty}d\Omega \Omega
\alpha^{2}_{ij}F^{ph,sf}(\Omega)/[(\omega_{n}-\omega_{m})^{2}+\Omega^{2}]. \]
$\Theta$ is the Heaviside function and $\omega_{c}$ is a cutoff
energy.  The quantities $\mu^{*}_{ij}(\omega\ped{c})$ are the elements of the $3\times 3$
Coulomb pseudopotential matrix. Finally,
$N^{\Delta}_{j}(i\omega_{m})=\Delta_{j}(i\omega_{m})/
{\sqrt{\omega^{2}_{m}+\Delta^{2}_{j}(i\omega_{m})}}$ and
$N^{Z}_{j}(i\omega_{m})=\omega_{m}/{\sqrt{\omega^{2}_{m}+\Delta^{2}_{j}(i\omega_{m})}}$.
The electron-boson coupling constants are defined as
$\lambda^{ph,sf}_{ij}=2\int_{0}^{+\infty}d\Omega\frac{\alpha^{2}_{ij}F^{ph,sf}(\Omega)}{\Omega}$.

To solve these equations (\ref{eq:EE1} and \ref{eq:EE2}) it is first necessary to specify a certain number of input parameters which depend from the particular characteristics in the studied system.
Many times it is possible through drastic approximations to reduce the number of input parameters, which are not always known, without renouncing to accurately describe the physics of the system.
In the case of a three bands model we have nine electron-phonon spectral functions $\alpha^{2}_{ij}F^{ph}(\Omega)$, nine electron-antiferromagnetic spin fluctuation spectral functions, $\alpha^{2}_{ij}F^{sf}(\Omega)$, nine elements of the Coulomb pseudopotential matrix $\mu_{ij}^{*}(\omega_{c})$, nine nonmagnetic $\Gamma^{N}\ped{ij}$ and nine paramagnetic $\Gamma^{M}\ped{ij}=0$ impurity-scattering rates.
Luckily, a lot of these parameters can be extracted from experiments and some can be fixed by suitable approximations.
In fact, fortunately, the systems we want to describe, the iron pinictides, has particular characteristics that allow numerous strong approximations aimed at reducing the number of free parameters. Despite this, the model still allows the main properties of these materials to be described in an extremely precise way.
In particular, we refer to experimental data taken on high quality samples, so we can rather safely assume a negligible disorder so we can put the scattering from non-magnetic and magnetic impurities $\Gamma^{N,M}\ped{ij}$ equal to zero. We know that in these materials the total electron-phonon coupling constant is small (the upper limit of these compounds is $\approx0.35$ \cite{Boeri}) and the phonons mainly provide \emph{intra}band coupling so that $\lambda^{ph}_{ij}\approx0$ \cite{Mazin1}. Furthermore it is well established that the superconducting glues is provided from antiferromagnetic spin fluctuations. These last bosons mainly provide \cite{Mazin1} \emph{interband coupling and just between hole and electron bands}, so that $\lambda^{sf}_{12}=\lambda^{sf}_{21}=\lambda^{sf}_{ii}=0$. For reducing the number of free parameters without alter the physics of the system we put, in first approximation, the phonon \emph{intra}band coupling equal to $0.1$ so that $\lambda^{ph}_{ii}=0.1$ and the Coulomb pseudopotential matrix \cite{Mazin3,Umma1,Umma2,Umma3} $\mu^{*}_{ii}(\omega_{c})=\mu^{*}_{ij}(\omega_{c})=0$. As we discussed before the maximum value of the total electron-phonon coupling is stimated less then $0.35$ a bit larger of the Coulomb pseudopotential that has opposite sign, so \textit{in the first approximation}, and for reducing the number of free parameters, we put the pseudopotential equal to zero and the intraband phonon coupling equal to $0.1$ because the second reduces the first. Of course this doesn't means that the phonons are absent but just that the final result in the calculus of a lot of physical properties is not influenced of the their presence. Within these approximations, the electron-boson coupling-constant matrix $\lambda_{ij}$ becomes:
\cite{Umma1,Umma2,Umma3}:
\begin{equation}
\vspace{2mm} %
\lambda_{ij}= \left (
\begin{array}{ccc}
  0.1                 &         0                 &               \lambda^{sf}_{13}            \\
  0                &               0.1               &                \lambda^{sf}_{23}            \\
  \lambda^{sf}_{31}=\lambda^{sf}_{13}\nu_{13} &   \lambda^{sf}_{32}=\lambda^{sf}_{23}\nu_{23}  & 0.1 \\
\end{array}
\right ) \label{eq:matrix}
\end{equation}
where $\nu_{ij}=N_{i}(0)/N_{j}(0)$, and $N_{i}(0)$ is the normal density of states at the Fermi level for the $i$-th band.
The coupling constants $\lambda_{ij}^{sf}$ are defined through the electron-antiferromagnetic spin fluctuation spectral functions (Eliashberg functions) $\alpha^2_{ij}F_{ij}^{sf}(\Omega)$. We choose these functions to have a Lorentzian shape \cite{Umma1,Umma2,Umma3} which reproduce the experimentally measured form quite well \cite{Inosov}:
\begin{equation}
\alpha_{ij}^2F^{sf}_{ij}(\Omega)= C_{ij}\big\{L(\Omega+\Omega_{ij},Y_{ij})-
L(\Omega-\Omega_{ij},Y_{ij})\big\},
\end{equation}
where
\[
L(\Omega\pm\Omega_{ij},Y_{ij})=\frac{1}{(\Omega \pm\Omega_{ij})^2+Y_{ij}^2}
\]
and $C_{ij}$ are normalization constants, necessary to obtain the proper values of $\lambda_{ij}$, while $\Omega_{ij}$ and $Y_{ij}$ are the peak energies and the half-widths of the Lorentzian functions, respectively \cite{Umma3}.  In all the calculations we set $\Omega_{ij}=\Omega_{0}$, i.e. we assume that the characteristic energy of antiferromagnetic spin fluctuations is a single quantity for all the coupling channels, and  $Y_{ij}= \Omega_{0}/2$, based on the results of inelastic neutron scattering measurements \cite{Inosov}.

The peak energy of the Eliashberg functions, $\Omega_0$, can be directly associated to the experimental critical temperature, $T_c$, by using the empirical law $\Omega_{0}=4.65k_{B}T_{c}$ that has been demonstrated to hold, at least approximately, for iron pnictides \cite{Paglione1,Paglione2}. With all these approximations, necessary to reduce the number of free parameters, this is the more simple model that can still grasp the essential physics of iron compounds.
The cut-off energy is $\omega_{c}=6.7568\Omega_{0}$. We assume, just for simplicity, that the electron phonon spectral functions have the same shape of the electron-antiferromagnetic spin fluctuation spectral functions.

The factors $\nu_{ij}=\frac{Ni(0)}{Nj(0)}$ that enter the definition of $\lambda_{ij}$ (eq. 3) are free parameters so we examine five different exaustive situations: first case $\nu_{13}=0.2$ and $\nu_{23}=1$, second case $\nu_{13}=0.5$ and $\nu_{23}=1$, third case $\nu_{13}=1$ and $\nu_{23}=1$, fourth case $\nu_{13}=2$ and $\nu_{23}=1$ and fifth case $\nu_{13}=5$ and $\nu_{23}=1$. At the end, fixed the $N_{i}(0)$, for each case, we have just two free parameters $\lambda_{13}$ and $\lambda_{23}$ so we change $\lambda_{13}$ and we fix $\lambda_{23}$ in order to obtain the correct critical temperature. In the known multiband superconductors and specifically in the iron pnictides the values of the densities of the states at the Fermi level $N_{i}(0)$ relating to the various bands are roughly of the same order of magnitude. Therefore in the five cases examined I have exhausted all the possible cases that have occurred to date. In principle it is easy to calculate the densities of the states at the Fermi level for the bands of a given material while it is much more complicated to calculate the electron boson coupling constants especially when the mechanism is the antiferromagnetic spin fluctuactions.
It is possible to define a total electron boson coupling constant (with sign)
$\lambda_{tot}=\sum_{i,j=1}^2 N_i(0)\lambda_{ij}/\sum_{i=1}^2 N_i(0)$ where the coupling constant related to antiferromagnetic spin fluctuactions are negative.
\section{Discussion}
In figure 1 (2) it is possible to see $|\lambda_{23}|$ ($\lambda_{tot}$) as a function of $|\lambda_{13}|$ in the various cases examined. The relevant thing is that these curves are universal, they are valid for any critical temperature. In the figure 1 it is possible to see that all curves pass for the point $|\lambda_{13}|=0.95$ and $|\lambda_{23}|=2.37$. With the same values of $|\lambda_{13}|$ and $|\lambda_{13}|=0.95$3 what changes is only the total value of the electron-boson coupling. Then in the particular case when $|\lambda_{13}|=0.95$ and $|\lambda_{23}|=2.37$ the variation range of the total electron-boson coupling is: $-2.2<\lambda_{tot}<-1.9$.
The universality derives from the fact that we impose a very strong (experimental) constraint on the energy of the peak of the spectral functions $\Omega_{0} = 4.65k_{B}T_{c}$. The universality of figures 1 and 2 lies in the fact that, once the densities of the states at the Fermi level $N_{i}(0)$ relating to the single bands have been fixed, there is an unequivocal relationship between the two coupling constants $\lambda_{13}$ and $\lambda_{23}$: once one is fixed, there can exist only one value of the other which reproduces the correct $T_{c}$.
From these curves it is also possible to see that $0<|\lambda_{23}|<3.5$ and $0<|\lambda_{13}|<7.5$ and also this result does'not depend from the particular critical temperature as that $1.1<|\lambda_{tot}|<2.3$. This means that, in principle, for all iron pnictides the total coupling is, in absolute value, less than 2.3 and this fact means that they can be just in a state of moderate strong coupling.
In figure 3 the $|\Delta_{i}|/k_{b}T_{c}$ ratios are shows for the three gaps with three different critical temperatures ($T_{c}=37$ K, $T_{c}=57$ K and $T_{c}=200$ K) and, as you can see, the results are perfectly superimposable.
 Here of course $|\Delta_{i}|$ is calculated from the solution of Eliashberg equations, at $T<<T_{c}$, by using Pad\`{e} approximants. The same happen also for the superconducting densities of states, as it is possible to see in figure 4. The superconducting densities of states, calculated at $T=T_{c}/12$, for $T_{c}=37$ K, $T_{c}=57$ K and $T_{c}=200$ K versus $\omega/\Omega_{0}$ in the first ($\nu_{13=0.2}$ and $\nu_{23=1}$ with $\lambda_{13}=6.0000$ and $\lambda_{23}=1.1577$) and third case ($\nu_{13}=1$ and $\nu_{23}=1$ with $\lambda_{13}=0.95$ and $\lambda_{23}=2.37$),
 are very different in the two cases but within each case, for different values of the critical temperature, they are perfectly superimposable. The first case with $\lambda_{13}=6$ can be considered as extreme because but in any case the scaling law continue to hold perfectly.
Finally we have tried to study what happens in the case of extreme strong coupling when the ratio $\frac{k_{B}T_{c}}{\Omega_{0}}$ is equal to one. We will study the third case ($\nu_{13}=\nu_{23}=1$).
The rate $\frac{k_{B}T_{c}}{\Omega_{0}}=1$ is considered extreme strong coupling and not physical because we find, as it is shown in figure 5, $\lambda_{tot}\geq 20$. For these values of the coupling constants it becomes problematic to define the value of the gap as well because the equation that defines it has more solutions \cite{Umma4}. Furthermore, in this regime it is probable that Migdal's theorem no longer holds and Eliashberg's equations become enormously more complicated. Obviously this situation has no connection with iron pnictides or any other known multiband systems.
\section{Conclusions}
In this article it has been shown that the three-band model has universal aspects as the link between $\lambda_{23}$ and $\lambda_{13}$ or the value of $|\Delta_{i}|/k_{b}T_{c}$ that are independent of the particular features of a given system and from the particular critical temperature.
These universal aspects are relate to the assumption that the typical bosonic energy is correlated to the critical temperature as shown by experimental data. By assuming $\Omega_{0} = 4.65k_{B}T_{c}$ a strict constraint is imposed to the value of the electron boson coupling constant. A similar conclusion may be derived from the analysis of the Allen-Dynes formula \cite{Allen} for the critical temperature in a one-band model. We here prove that in a fully numerical solution of the Eliashberg equation for a multi-band model such a constraint hold with great accuracy.
\section{ACKNOWLEDGMENTS}
G.A.U. acknowledges support from the MEPhI Academic Excellence Project (Contract No. 02.a03.21.0005).\\

\begin{figure}[ht]
\begin{center}
\includegraphics[keepaspectratio, width=\columnwidth]{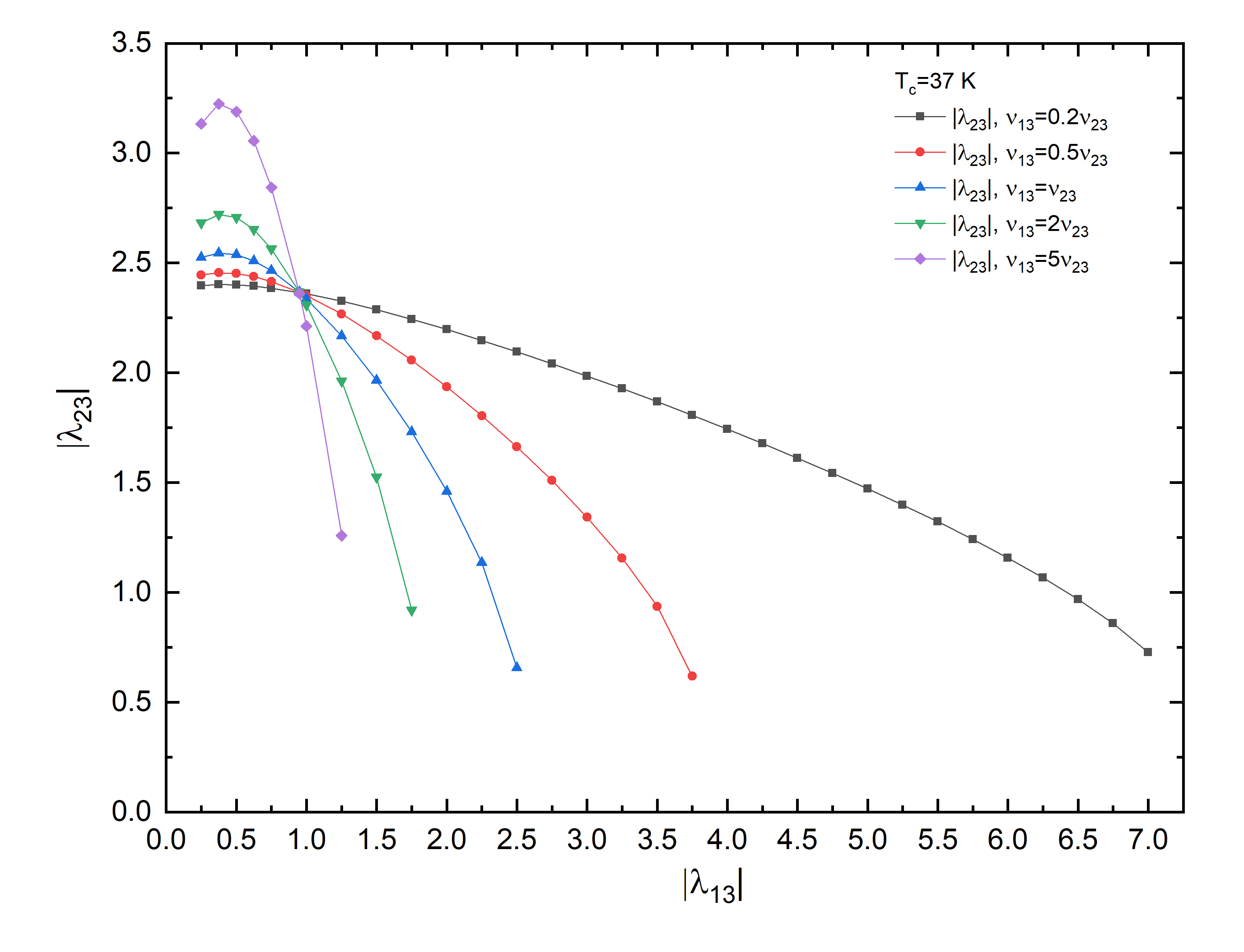}
\caption{(Color online) $|\lambda_{23}|$ versus $|\lambda_{13}|$.}
\end{center}
\end{figure}

\newpage

\begin{figure}[ht]
\begin{center}
\includegraphics[keepaspectratio, width=\columnwidth]{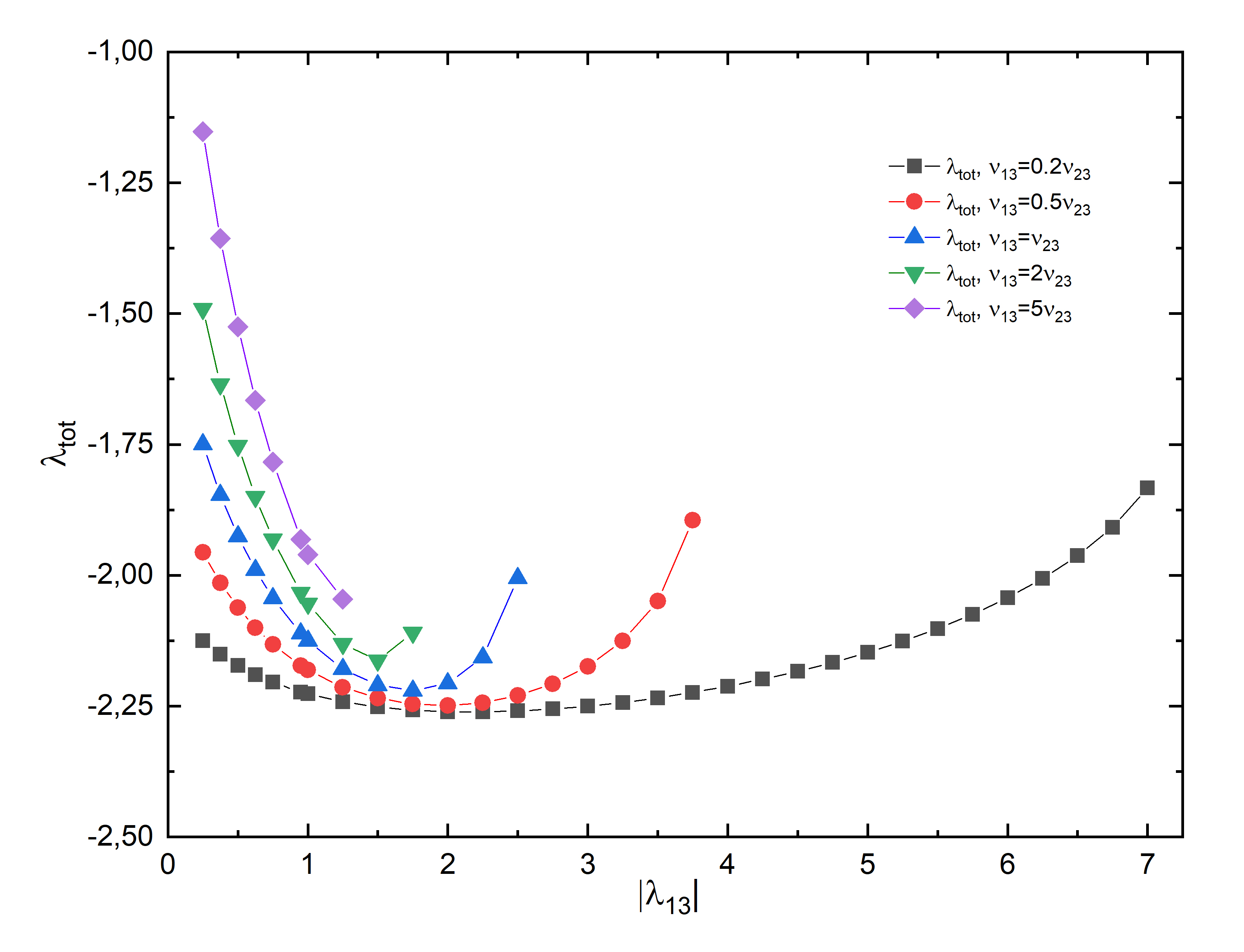}
\caption{(Color online) $\lambda_{tot}$ versus $|\lambda_{13}|$.}
\end{center}
\end{figure}

\newpage
\begin{figure}[ht]
\begin{center}
\includegraphics[keepaspectratio, width=\columnwidth]{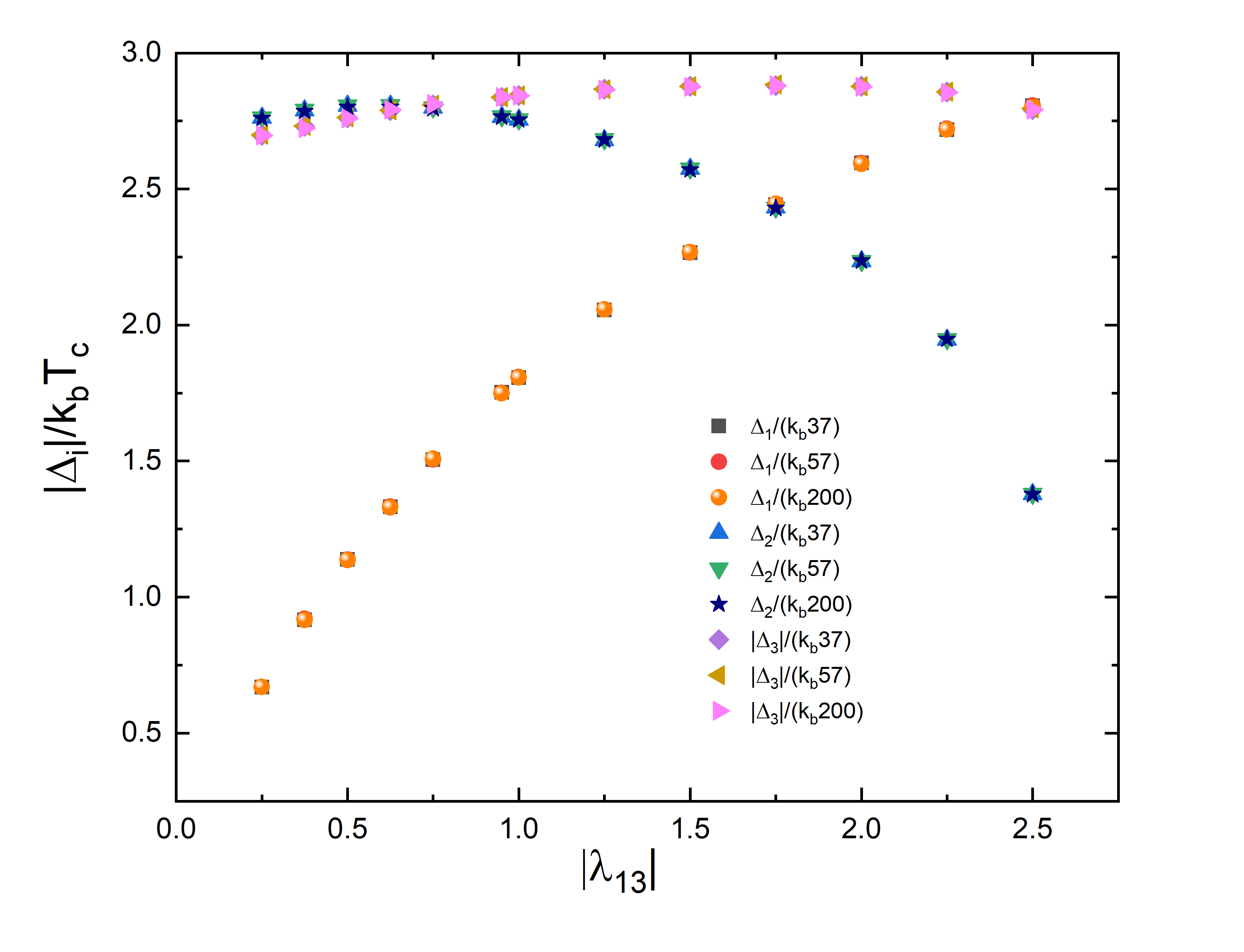}
\caption{(Color online) $|\Delta_{i}|/k_{B}T_{c}$ for $T_{c}=37$ K, $T_{c}=57$ K and $T_{c}=20$ K versus $|\lambda_{13}|$ in the case where the values of the partial dos at the Fermi level ($N_{i}(0)$) are all equals ($\nu_{13}=\nu_{23}=1$).}
\end{center}
\end{figure}

\newpage
\begin{figure}[ht]
\begin{center}
\includegraphics[keepaspectratio, width=\columnwidth]{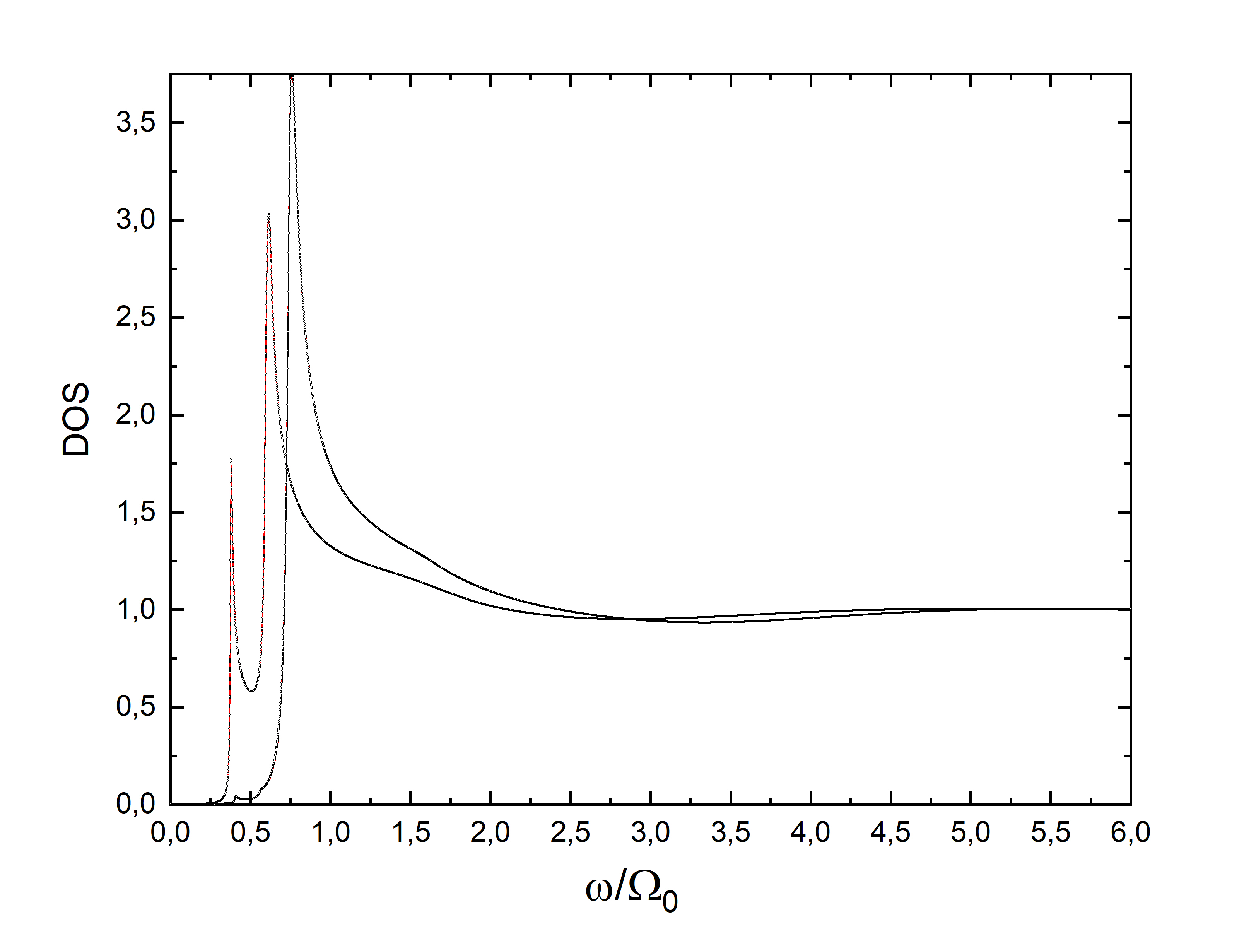}
\caption{(Color online) Densities of states, calculated at $T=T_{c}/12$, for $T_{c}=37$ K (red line), $T_{c}=57$ K (black line) and $T_{c}=200$ K (open black circles) versus $\omega/\Omega_{0}$ in the first ($\nu_{13}=0.2$ and $\nu_{23}=1$ with $\lambda_{13}=6.0000$ and $\lambda_{23}=1.1577$) and third case ($\nu_{13}=1$ and $\nu_{23}=1$ with $\lambda_{13}=0.9500$ and $\lambda_{23}=2.3657$).}
\end{center}
\end{figure}

\newpage
\begin{figure}[ht]
\begin{center}
\includegraphics[keepaspectratio, width=\columnwidth]{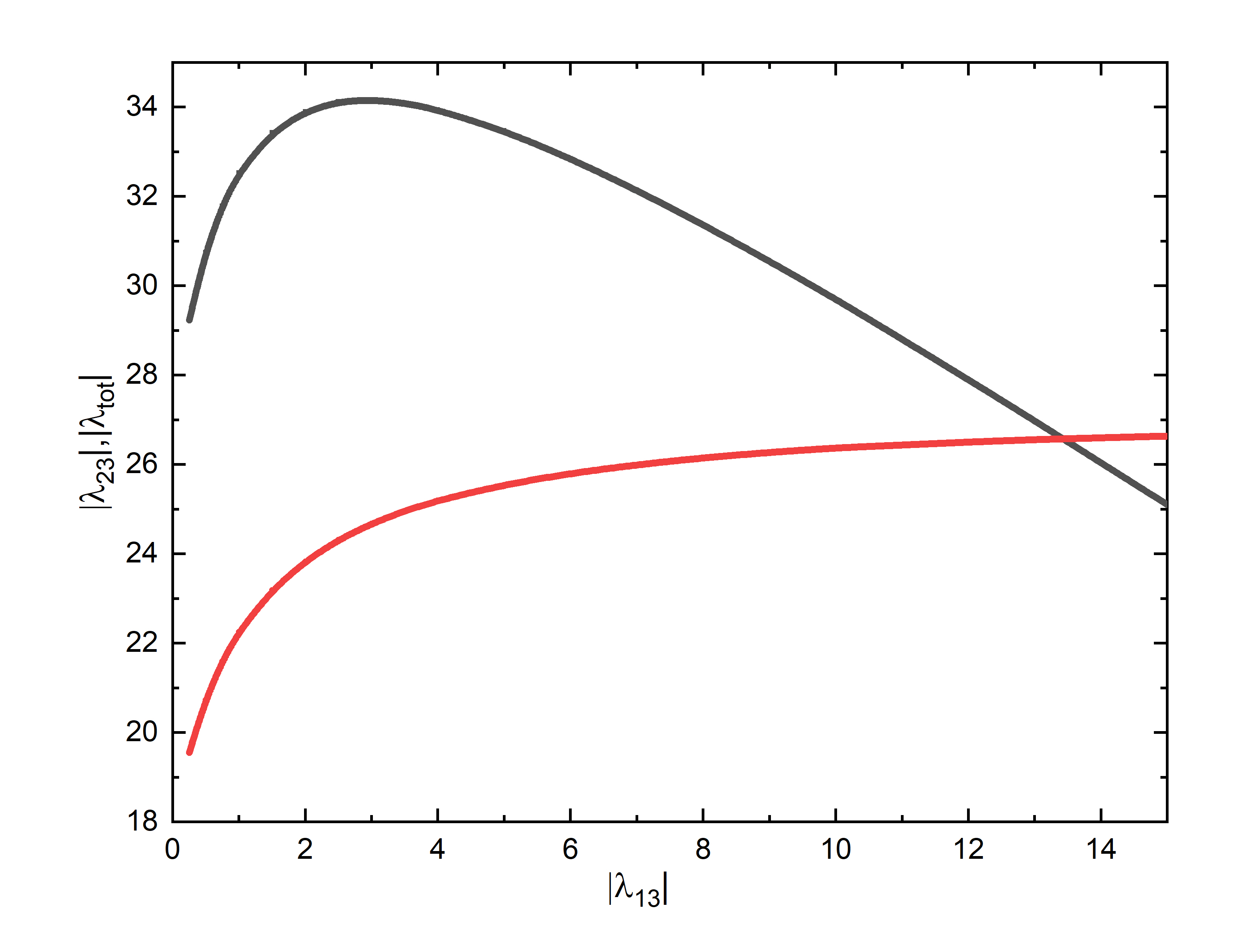}
\caption{(Color online) $\lambda_{tot}$ versus $|\lambda_{13}|$ (solid red line) and  $\lambda_{23}$ versus $|\lambda_{l3}|$ (black solid line) in the extreme strong coupling case ($\frac{k_{B}T_{c}}{\Omega_{0}}=1$) when the values of the partial dos at the Fermi level ($N_{i}(0)$) are all equals ($\nu_{13}=\nu_{23}=1$).}
\end{center}
\end{figure}

\newpage
\end{document}